\documentclass[conference]{IEEEtran}
\IEEEoverridecommandlockouts
\usepackage{amsmath}
\usepackage{balance}
\usepackage{cite}
\usepackage{acronym}
\usepackage{subfigure}
\usepackage{hyperref}
\usepackage{epsf}
\usepackage{graphicx}
\usepackage{flushend}

\acrodef{SC-FDMA}[SC-FDMA]{single-carrier frequency-domain multiple-access}
\acrodef{LOS}[LOS]{line-of-sight}
\acrodef{NLOS}[NLOS]{non-line-of-sight}
\acrodef{SINR}[SINR]{signal-to-interference-and-noise ratio}
\acrodef{SNR}[SNR]{signal-to-noise ratio}
\acrodef{FH}[FH]{frequency hopping}

\begin{document}

\title
{Performance Analysis of Fifth-Generation \\ Cellular Uplink}
\author{\IEEEauthorblockN{ Don Torrieri,\IEEEauthorrefmark{1}
Salvatore Talarico,\IEEEauthorrefmark{2}
and Matthew C. Valenti\IEEEauthorrefmark{2} } \IEEEauthorblockA
{\IEEEauthorrefmark{1}%
U.S. Army Research Laboratory, Adelphi, MD, USA \\ \IEEEauthorrefmark
{2}West Virginia University, Morgantown, WV, USA} }%
\maketitle

\begin{abstract}
Fifth-generation cellular networks are expected to exhibit at least three
primary physical-layer differences relative to fourth-generation ones:
millimeter-wave propagation, antenna-array directionality, and densification
of base stations. In this paper, the effects of these differences on the
performance of single-carrier frequency-domain multiple-access uplink systems
with frequency hopping are assessed. A new analysis, which is much more
detailed than any other in the existing literature and accommodates actual
base-station topologies, captures the primary features of uplink
communications. Distance-dependent power-law, shadowing, and fading models
based on millimeter-wave measurements are introduced. The beneficial effects
of base-station densification, highly directional sectorization, and frequency
hopping are illustrated.

\end{abstract}

\section{Introduction}

This paper provides a performance analysis of the uplink of a fifth-generation
cellular network in its primary mode of operation. It is assumed that
millimeter-wave frequencies will be adopted, but that the basic structure of
the fourth-generation (4G) \ac{SC-FDMA} uplink systems will be maintained. The first assumption is widely
supported in the current literature on fifth-generation (5G) systems \cite{andr},
\cite{wang}. The second one is based on the favorable characteristics of
\ac{SC-FDMA} and the critical importance of transmitter power efficiency, which is
enabled by the low peak-to-average power ratio of \ac{SC-FDMA} and similar
single-channel modulations \cite{deb}, \cite{ali}. It is further assumed that 5G systems will exploit frequency hopping, as is done in 4G/LTE systems \cite{dahlman:2013}.

The analysis in this paper applies the methodology of \cite{torval1}, which we
call deterministic geometry. Unlike stochastic geometry, deterministic
geometry can accommodate arbitrary topologies with distance-dependent
propagation models. The distance dependence of the propagation and fading
models accounts for the fact that mobiles close to the base station have a
\ac{LOS} path, but the more distant mobiles do not.

The\ \emph{conditional} outage probability of a \ac{SC-FDMA} uplink is derived,
where the conditioning is with respect to an arbitrary network topology. In
the numerical examples, we use an actual deployment of the base stations. Each
simulation trial generates a network realization in which the mobile
placements are drawn from the uniform clustering distribution with each mobile
having an exclusion zone \cite{torval1}. For each realization of the network,
the outage probability is computed for a reference link. By averaging over
many network realizations, the average outage probability and other
statistical performance measures are computed.

The remainder of the paper is organized as follows. Section
\ref{sec:Model} describes the system model, which accounts for distance dependent fading, path loss, and shadowing based on millimiter-wave measurements, and for highly directional antennas and sectorization. Section III provides a closed-form expression for the conditional outage probability.  Section  IV  provides a numerical evaluation of a 5G system. Finally, the paper concludes in Section V.

\section{Network Model}\label{sec:Model}

\begin{figure}[tb]%
\centering
\includegraphics[width=9cm]{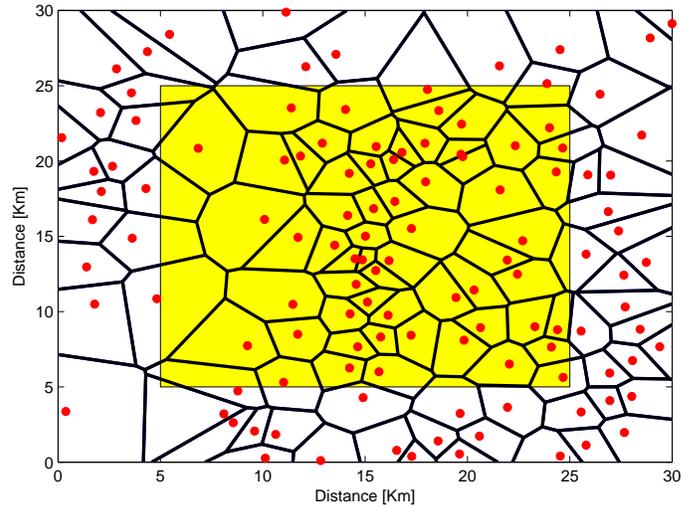}
\caption{Actual base station locations from a current cellular deployment.
Base stations are represented by large circles, and Voronoi cell boundaries
are represented by thick lines.} \label{Fig.1}
\end{figure}

In the network model, $C$ base stations and $M$ mobiles are confined to a
finite area. As an example, Fig. \ref{Fig.1} depicts an actual deployment of $C=121$
base stations extracted from a database of base-station locations in the
United Kingdom. Base stations are represented by large circles, and Voronoi
cell boundaries are represented by thick lines. The network occupies $900$
km$^{2}$ inside a square.

For practical reasons such as the high propagation loss and the complexity of
channel state estimation at millimeter-wave frequencies, the antenna arrays at
the base stations and mobiles exploit beamforming rather than spatial
multiplexing for uplink transmissions \cite{sun}. Densification, high
mobility, and the severe impact of blockages at millimeter-wave frequencies
cause frequent handoffs and hence the need for rapid beam alignments.
Sectorization, which is the division of base-station coverage into $\zeta$
fixed angular sector beams centered at the base station, is used to reduce
beam-alignment delays and pilot contamination. At millimeter-wave frequencies,
the beams can be implemented using many antenna elements, perhaps hundreds,
and hence have narrow beamwidths and very small sidelobes and backlobes. Each
mobile transmits through an adaptive and highly directional antenna array.

The scalar $S_{l},$ $l=1,2,...,\zeta C,$ represents the $l$th sector or its
receiver, and the scalar $X_{i},$ $i=1,2,...,M,$ represents the $i$th mobile.
The vector $\mathbf{S}_{l},$ $l=1,2,...,\zeta C,$ represents the location of
the $l$th sector receiver, and the vector $\mathbf{X}_{i},$ $i=1,2,...,M,$
represents the location of the $i$th mobile. The \textit{normalized sector
beam pattern} associated with $S_{l}$ is%
\begin{equation}
B_{l}\left(  \theta\right)  =\left\{
\begin{array}
[c]{cc}%
1, & \psi_{l}\leq\theta\leq\psi_{l}+2\pi/\zeta\\
b, & otherwise
\end{array}
\right.
\end{equation}
where $\psi_{l}$ is the offset angle of the beam pattern, and $b$ is the
relative sidelobe and backlobe level.

Let ${\mathcal{X}}_{l}\ $denote the set of mobiles served by sector $S_{l}$.
Let $X_{r}\in\mathcal{X}_{j}$ denote a reference mobile that transmits a
desired signal to a reference receiver $S_{j}$. Let $\mathsf{g}(i)$ denote a
function that returns the index of the sector serving $X_{i}$ so that
$X_{i}\in{\mathcal{X}}_{l}$ if $\mathsf{g}\left(  i\right)  =l$. The sector
$S_{\mathsf{g}\left(  i\right)  }$ that serves mobile $X_{i}$ is assumed to be
the one with minimum local-mean path loss when the mainlobe of the transmit
beam of $X_{i}$ is aligned with the sector beam of $S_{\mathsf{g}\left(
i\right)  }.$ Thus, the serving sector has index
\begin{equation}
\mathsf{g}\left(  i\right)  =\mathrm{\arg\max}_{l}\,\left\{  10^{\xi_{i,l}%
/10}f\left(  ||\mathbf{S}_{l}-\mathbf{X}_{i}||\right)  ,\text{ }X_{i}%
\in\mathcal{A}_{l}\right\}
\end{equation}
where $\xi_{i,l}$ is a \textit{shadowing factor }for the link from $X_{i}$ to
$S_{l}$, $f\left(  \cdot\right)  $ is the area-mean path-loss function,
$||\mathbf{\cdot}||$\ is the Euclidean norm, and $\mathcal{A}_{l}$ denote the
set of mobiles \emph{covered} by the sector beam of $S_{l}$. In the absence of
shadowing, the serving sector will be the receiver that is closest to $X_{i}$.
In the presence of shadowing, a mobile may actually be associated with a
sector that is more distant than the closest one if the shadowing conditions
are sufficiently better.

The area-mean path-loss function is a function of the distance $d$ between a
source and destination and is expressed as the attenuation power law
\begin{equation}
f\left(  d\right)  =\left(  \frac{d}{d_{0}}\right)  ^{-\alpha\left(  d\right)
},\text{ \ }d\geq d_{0}\label{eqn:pathloss}%
\end{equation}
where $\alpha\left(  d\right)  $ is the attenuation power-law exponent, and
$d_{0}$ is a reference distance that is less than or equal to the minimum of
the near-field radius and the exclusion-zone radius of a mobile.

The distance-dependent model of the power-law exponent for millimeter-wave
frequencies takes into account the area-mean attenuation due to blockages and
reflections that occur over network links. This model reflects the empirical
fact that $\alpha\left(  d\right)  $ differs substantially for \ac{LOS} and
\ac{NLOS} links, tending toward $\alpha_{\min}$ for the usually
shorter LOS links and tending toward a much larger $\alpha_{\max}$ for the
usually longer \ac{NLOS} links \cite{rang}, \cite{akd}, \cite{rapp}. Empirical data
indicates that usually there is a small range of link lengths for which there
are both \ac{LOS} and \ac{NLOS} links. Therefore, $\alpha\left(  d\right)  $ is modeled
as a monotonically increasing function:
\begin{equation}
\alpha\left(  d\right)  =\alpha_{\min}+\left(  \alpha_{\max}-\alpha_{\min
}\right)  \tanh\left(  \mu d\right)
\end{equation}
which indicates that $\alpha_{\min}\leq\alpha\left(  d\right)  <\alpha_{\max
}.$ The parameter $\mu$ controls the transition rate from $\alpha_{\min}$ to a
value close to $\alpha_{\max}$.

Local-mean large-scale terrain effects are included in the shadowing model.
The shadowing factor can be derived from a deterministic terrain model or can
be modeled as a random variable in a statistical model. In this paper, we
assume lognormal shadowing in which the $\{\xi_{i,j}\}$ are independent,
identically distributed, zero-mean Gaussian random variables with a
distance-dependent variance.

For millimeter-wave frequencies, empirical data \cite{rang}, \cite{akd},
\cite{rapp} indicates that the effect of the shadowing increases for the
usually longer \ac{NLOS} links. Since there is a small range of link lengths for
which there are both \ac{LOS} and \ac{NLOS} links, the standard deviation of the
shadowing factor for millimeter-wave frequencies is modeled as a monotonically
increasing function:%
\begin{equation}
\sigma_{s}\left(  d\right)  =\sigma_{\min}+(\sigma_{\max}-\sigma_{\min}%
)\tanh\left(  \mu d\right)
\end{equation}
which indicates that $\sigma_{\min}\leq\sigma_{s}\left(  d\right)
<\sigma_{\max}.$

The fading is assumed to have a Nakagami distribution function. Since the
fading becomes more severe for the longer links, the distance-dependent
Nakagami parameter is modeled as a monotonically decreasing function:%
\begin{equation}
m\left(  d\right)  =m_{\max}-(m_{\max}-m_{\min})\tanh\left(  \mu d\right)
\label{fade}%
\end{equation}
which indicates that $m_{\min}\leq m\left(  d\right)  < m_{\max}.$

Frequency hopping \cite{torr} is used in \ac{SC-FDMA} uplink systems to provide the
diversity that will mitigate the effects of frequency-selective fading and
intersector interference. Because of network synchronization and similar
propagation delays for the mobiles associated with a cell sector, synchronous
orthogonal frequency-hopping patterns can be allocated so that at any given
instant in time, there is no \emph{intrasector} interference. The
frequency-hopping patterns transmitted by mobiles in other sectors are not
generally orthogonal to the patterns in a reference sector, and hence produce
\emph{intersector} interference. The varying propagation delays from the
interfering mobiles cause their frequency-hopping signals to be asynchronous
with respect to the desired signal. Duplexing prevents uplink interference
from downlink transmissions.

Each mobile uses a frequency-hopping pattern over a hopset with $L$ disjoint
frequency channels. Let $L_{l},$ $l=1,2,\ldots,\zeta C,$ denote positive
integer divisors of $L$ such that $L/L_{l}\geq2.$ Each mobile in
${\mathcal{X}}_{l}\ $is assigned a distinct block of $L_{l}$ contiguous
frequency channels during each of its hop intervals, and the block may change
to any of $L/L_{l}$ disjoint spectral regions with every hop. Consider an
uplink \emph{reference signal} that traverses a reference link from a
reference mobile $X_{r}$ to a reference receiver $S_{j}.$ Because of a
possible incomplete spectral overlap, the received interference power from
mobile $X_{i}$ at $S_{j}$ when the mobile's signal collides with the reference
signal is reduced by the \textit{spectral factor}
\begin{equation}
F_{l}=\min\left(  L_{j}/L_{l},1\right)  .
\end{equation}

Each active mobile points its antenna beam toward its base station. The beam
pattern is modeled with two gains: one for the mainlobe and another for the
sidelobes and backlobes. The \emph{normalized beam gain} at the reference
receiver $S_{j}$ due to the angular offset of this beam pattern, is defined
as
\begin{equation}
b_{i,j}=\left\{
\begin{array}
[c]{cc}%
1, & \frac{\left(  \mathbf{S}_{j}-\mathbf{X}_{i}\right)  ^{T}\left(
\mathbf{S}_{\mathsf{g}\left(  i\right)  }-\mathbf{X}_{i}\right)
}{||\mathbf{S}_{j}-\mathbf{X}_{i}||||\mathbf{S}_{\mathsf{g}\left(  i\right)
}-\mathbf{X}_{i}||}>\cos\left(  \frac{\Theta}{2}\right)  \\
a, & otherwise
\end{array}
\right.
\end{equation}
where $\Theta$ is the beamwidth of the antenna mainlobe, superscript $T$
denotes the transpose, and $a$ is the sidelobe and backlobe level$.$

\begin{figure}[tb]%
\centering
\includegraphics[width=9cm ]{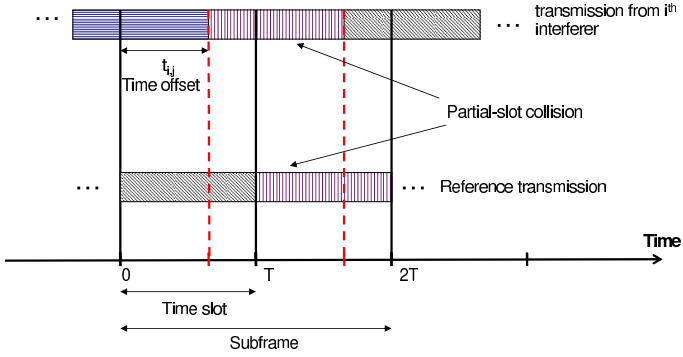}
\caption{Illustration of co-channel interference at base station $S_{j}$ produced by an interfering mobile $X_i$ arriving at the base station with a relative timing offset of $t_{i,j}$.  Each block in the illustration represents a hop, and the selected channel is indicated by the shading of the block.  A partial-slot collision is indicated, where the interfering mobile has selected the channels used by the reference mobile during the second slot of the subframe.  However, due to the asynchronism, the collision only extends for part of the slot.} \label{Fig.2}
\end{figure}

Associated with each potentially interfering mobile is a hop transition time $t_{i,j}$ at $S_{j}$ relative to the hop transition time of a pair of hop intervals of the reference signal.  Fig. \ref{Fig.2} illustrates the relative timing between the reference mobile and the $i^{th}$ interfering mobile.   The reference mobile transmits a turbo codeword of duration $2T$, which is aligned with the subframe.  It is assumed that the frequency
separation of the two frequency channels of the two slots is sufficient for
independent fading of fixed amplitude in each slot. If the base stations and mobiles are synchronized, then%
\begin{equation}
t_{i,j}=[(||\mathbf{S}_{j}-\mathbf{X}_{r}||-||\mathbf{S}_{j}-\mathbf{X}%
_{i}||)/c]\operatorname{mod}T
\end{equation}
where $c$ is the speed of an electromagnetic wave.
As illustrated in Fig. \ref{Fig.2}, the reference signal
encounters four time periods of potential interference from an active mobile
$X_{i}$: $0\leq t\leq t_{i,j},$ $t_{i,j}\leq t\leq T,$ $T\leq t\leq
t_{i,j}+T,$ and $t_{i,j}+T\leq t\leq2T$. The generic index $k\in\{1,2,3,4\}$,
denotes a time period of the subframe with duration that varies with each
$t_{i,j}.$ In the example provided in Fig. \ref{Fig.2}, co-channel interference occurs during the third time period, where the interfering mobile has selected the channels used by the
reference mobile. The \textit{fractional duration} of the each of the four subframe time periods relative to the subframe period $2T$ are
\begin{equation}
C_{i,j,k}=\left\{
\begin{array}
[c]{cc}%
\frac{t_{i,j}}{2T}, & k=1,3\\
\frac{T-t_{i,j}}{2T}, & k=2,4
\end{array}
\right.  .
\end{equation}

The set of indices of potentially interfering mobiles is $\mathcal{S}=\left\{
i:X_{i}\notin{\mathcal{X}}_{j}\right\}  $. Let $N_{l}$ denote the number of
mobiles associated with sector $S_{l}.$ Because of the required orthogonality
of frequency blocks assigned to mobiles within each sector, $N_{l}\leq$
$L/L_{l}$ and any additional mobiles within the sector are reassigned to other
sectors. In view of the potential spectral overlaps, the maximum number of
interfering mobiles within a sector during a subframe time period is
$\min[\max\left(  L_{j}/L_{l},1\right)  ,N_{l}].$ Let $\mathcal{S}_{k}\subset$
$\mathcal{S}$ denote the set of interfering mobiles during subframe time
period $k.$ If $N_{l}\leq\max\left(  L_{j}/L_{l},1\right)  ,$ then all $N_{l}$
mobiles in sector $l$ are in $\mathcal{S}_{k}$. If $N_{l}>\max\left(
L_{j}/L_{l},1\right)  ,$ then some of the mobiles in sector $l$ cannot cause
interference during subframe time period $k$. In that case, we approximate by
randomly selecting a subset of the $N_{l}$ mobiles to be included in
$\mathcal{S}_{k}.$

Let $q_{i,k}$ denote the probability that the signal from a potentially
interfering mobile collides with the reference signal during subframe time
period $k$, $1\leq k\leq4.$ The \emph{activity\ probability} $p_{i}$ is the
probability that mobile $X_{i}$ is transmitting during time interval $[0,2T)$.
Assuming uniformly distributed frequency-hopping patterns that are orthogonal
within each sector,
\begin{equation}
q_{i,k}=\frac{\max\left(  N_{\mathsf{g}\left(  i\right)  }L_{\mathsf{g}\left(
i\right)  },L_{j}\right)  }{L}p_{i},\text{ \ }i\in\mathcal{S}_{k},\text{
\ }1\leq k\leq4
\end{equation}
and $q_{i,k}=0,$ otherwise$.$\

The instantaneous \ac{SINR} at sector
receiver $S_{j}$ when the desired signal is from $X_{r}\in\mathcal{X}_{j}$
fluctuates because potentially interfering signals do not always coincide with
the reference signal in time or frequency. Pilot sequences are used to
estimate the complex fading amplitudes in the receiver. Therefore, the
performance of the reference receiver is primarily a function of the
\emph{average \ac{SINR}} defined as the ratio of the average power of the signal to
the average power of the noise and interference, where the average is over the
time interval of a subframe and turbo codeword. Thus, the average \ac{SINR} during
a subframe is
\begin{equation}
\gamma_{r,j}=\frac{\overline{\rho}_{r,j}}{{\mathcal{N}}+\sum_{k=1}^{4}%
\sum_{i\in\mathcal{S}_{k}}I_{i,k}\rho_{i,j,k}C_{i,j,k}} \label{SINR1}%
\end{equation}
where $\mathcal{N}$ is the noise power, $\overline{\rho}_{r,j}$ is the average
received power from reference mobile $X_{r},$ and $\rho_{i,j,k}$ is the
received power from an interference signal that collides with the reference
signal during subframe time period $k.$ The indicators $I_{i,k}$ are Bernoulli
random variables with probabilities%
\begin{align}
P[I_{i,k}  &  =1]=q_{i,k},\text{ }P[I_{i,k}=0]=1-q_{i,k}\nonumber\\
\text{\ }i  &  \in\mathcal{S}_{k},\text{ \ }1\leq k\leq4.
\end{align}

Let $g_{i,j,k}$ denote the fading gain of the signal from mobile $X_{i}$ at
$S_{j}$ during time interval $k$. Assuming that the bandwidths of the
$L/L_{l}$ and $L/L_{j}$ disjoint spectral regions exceed the coherence
bandwidth, the \{$g_{i,j,k}\}$ are independent for each hop with
unit-mean, and $g_{i,j,k}=a_{i,j,k}^{2}$, where $a_{i,j,k}$ has a Nakagami
distribution with distance-dependent parameter $m_{i,j}$. Let ${P}_{i}$ denote
the maximum power from $X_{i}$ that could be received at the reference
distance $d_{0}$ in the absence of fading and shadowing. Allowing for the
spectral and beam factors, the received power from $X_{i}$ at $S_{j},$
$i\in\mathcal{S}_{k},$ during time interval $k$ is%
\begin{align}
\rho_{i,j,k} &  ={P}_{i}g_{i,j,k}10^{\xi_{i,j}/10}f\left(  ||\mathbf{S}%
_{j}-\mathbf{X}_{i}||\right)  b_{i,j}F_{\mathsf{g}\left(  i\right)  }%
B_{j}\left(  \theta_{i,j}\right)  \nonumber\\
\text{\ }i &  \in\mathcal{S}_{k},\text{ \ }1\leq k\leq4\label{pi}%
\end{align}
where $\theta_{i,j}$ is the arrival angle at $S_{j}$ of a signal from $X_{i},$
and $B_{j}\left(  \theta_{i,j}\right)  $ is the gain of the uplink beam
pattern of $S_{j}$.

Let $g_{r,j,1}$ and $g_{r,j,2}$ denote the unit-mean power gains due to the
independent fading of the frequency-hopping reference signal in subframe slots
1 and 2, respectively. The power gain of independent Nakagami fading with parameter
$m_{0}=m_{r,j}\ $in each slot has the gamma density function:
\begin{equation}
f_{r,j}(x)=\frac{m_{0}^{m_{0}}x^{m_{0}-1}\exp\left(  -m_{0}x\right)  }%
{\Gamma\left(  m_{0}\right)  }u\left(  x\right)  \label{2}%
\end{equation}
where $u\left(  x\right)  =1,$ $x\geq0,$ and $u\left(  x\right)  =0,$
otherwise. The average power gain due to fading is $\overline{g}_{r,j}=\left(
g_{r,j,1}+g_{r,j,2}\right)  /2.$ Using (\ref{2}), we obtain the density
function of $\overline{g}_{r,j}:$%
\begin{equation}
f_{0}(x)=\frac{\left(  2m_{0}\right)  ^{2m_{0}}x^{2m_{0}-1}\exp\left(
-2m_{0}x\right)  }{\Gamma\left(  2m_{0}\right)  }u\left(  x\right)
\end{equation}
which is the power of a Nakagami random variable with parameter $2m_{0}$. This
doubling of the Nakagami parameter indicates the beneficial effect of the
frequency hopping in mitigating frequency-selective fading. The average
received power from reference mobile $X_{r}$ is%
\begin{equation}
\overline{\rho}_{r,j}={P}_{r}\overline{g}_{r,j}10^{\xi_{r,j}/10}f\left(
d_{r}\right)  \label{pa}%
\end{equation}
where $d_{r}=||\mathbf{S}_{j}-\mathbf{X}_{r}||$ is the length of the reference link.

Power control ensures that the local-mean power received from each mobile in a
sector is equal to a constant. If this constant is the same for all sectors,
then
\begin{align}
P_{r}10^{\xi_{r,j}/10}f\left(  d_{r}\right)   &  =P_{i}10^{\xi_{i,\mathsf{g}%
\left(  i\right)  }/10}f\left(  ||\mathbf{S}_{\mathsf{g}\left(  i\right)
}-\mathbf{X}_{i}||\right)  ,\nonumber\\
\text{ \ }i  &  \in\mathcal{S}_{k}. \label{pc}%
\end{align}
Substituting (\ref{pi}), (\ref{pa}), and (\ref{pc}) into (\ref{SINR1}), we
obtain%
\begin{equation}
\gamma_{r,j}=\frac{\overline{g}_{r,j}}{\Gamma_{0}^{-1}+\sum_{k=1}^{4}%
\sum_{i\in\mathcal{S}_{k}}I_{i,k}\Omega_{i,j}g_{i,j,k}C_{i,j,k}}%
\end{equation}
where
\begin{equation}
\Omega_{i,j}=b_{i,j}F_{\mathsf{g}\left(  i\right)  }B_{j}\left(  \theta
_{i,j}\right)  \frac{10^{[\xi_{i,j}-\xi_{i,\mathsf{g}\left(  i\right)  }%
]/10}f\left(  ||\mathbf{S}_{j}-\mathbf{X}_{i}||\right)  }{f\left(
||\mathbf{S}_{\mathsf{g}\left(  i\right)  }-\mathbf{X}_{i}||\right)  }%
\end{equation}
is the ratio of the interference power from $X_{i}$ to the reference-signal
power, and%
\begin{equation}
\Gamma_{0}=\frac{P_{r}}{\mathcal{N}}10^{\xi_{r,j}/10}f\left(  d_{r}\right)
\label{gam}%
\end{equation}
is the \ac{SNR} at the sector receiver when the fading is absent.

\section{Outage Probability}

Let $\beta$ denote the minimum average \ac{SINR} required for reliable reception of
a signal from $X_{r}$ at its serving sector receiver $S_{j},$ $j=\mathbf{g}%
(r)$. An \emph{outage} occurs when the average \ac{SINR} of a signal from $X_{r}$
falls below $\beta$. The value of $\beta$ sets a limit on the code-rate $R$ of
the uplink, which is expressed in units of bits per channel use (bpcu), and
depends on the modulation and coding schemes, and the overhead losses due to
pilots, cyclic prefixes, and equalization methods. The exact dependence of $R$
on $\beta$ can be determined empirically through tests or simulation.

The set of $\{\Omega_{i,j}\},$ $i\in\mathcal{S}_{k},$ for reference receiver
$S_{j}$ is represented by the vector $\boldsymbol{\Omega}_{j}$. Conditioning
on $\boldsymbol{\Omega}_{j}$, the \emph{outage probability} of a desired
signal from $X_{r}\in{\mathcal{X}}_{j}$ that arrives at $S_{j}$ is
\begin{equation}
\epsilon=P\left[  \gamma_{r,j}\leq\beta\big|\boldsymbol{\Omega}_{j}\right]  .
\end{equation}
Because it is conditioned on $\boldsymbol{\Omega}_{j}$, the outage probability
depends on the particular network realization, which has dynamics over
timescales that are much slower than the fading or frequency hopping. We
define
\begin{equation}
\beta_{0}=2\beta m_{0},\text{ \ }z=\Gamma_{0}^{-1}%
\end{equation}
where the Nakagami parameter $m_{0}= \left\lfloor m_{r,j}\right\rfloor$ for the reference uplink signal
is assumed to be a positive integer. A derivation similar to the one in
\cite{torval1} yields
\begin{equation}
\epsilon=1-e^{-\beta_{0}z}\sum_{s=0}^{2m_{0}-1}{\left(  \beta_{0}z\right)
}^{s}\sum_{t=0}^{s}\frac{z^{-t}}{(s-t)!}H_{t}(\boldsymbol{\Omega})\text{ }
\label{a1}%
\end{equation}
where
\begin{equation}
H_{t}(\boldsymbol{\Omega})=
{\displaystyle\sum\limits_{\substack{\ell_{ik}\geq0\\\sum_{k=1}^{4}\sum
_{i\in\mathcal{S}_{k}}\ell_{ik}=t}}}
\prod_{k=1}^{4}\prod_{i\in\mathcal{S}_{k}}G_{\ell_{ik}}(i,j,k) \label{a2}%
\end{equation}
the summations in (\ref{a2}) are over all sets of nonnegative indices that sum
to $t$,
\begin{equation}
G_{\ell}(i,j,k)=%
\begin{cases}
1-q_{i,k}(1-\Psi_{i,j,k}^{m_{i,j}}), & \ell=0\\
\frac{q_{i,k}\Gamma(\ell+m_{i,j})}{\ell!\Gamma(m_{i,j})}\left(  \frac
{\Omega_{i,j}C_{i,j,k}}{m_{i,j}}\right)  ^{\ell}\Psi_{i,j,k}^{m_{i,j}+\ell} &
\ell>0
\end{cases}
\end{equation}
and
\begin{equation}
\Psi_{i,j,k}=\left(  \beta_{0}\frac{\Omega_{i,j}C_{i,j,k}}{m_{i,j}}+1\right)
^{-1}\hspace{-0.5cm},\text{ \ }i\in\mathcal{S}_{k},\text{ \ }1\leq k\leq4.
\label{a4}%
\end{equation}

\section{Numerical Results}

In the following examples, performance metrics are calculated by using a Monte
Carlo approach with $N$ simulation trials. In each simulation trial, a
realization of the network of Fig. 1 is obtained by placing $M$ mobiles within
it according to a uniform clustering distribution with each mobile having an
exclusion-zone radius set equal to $d_{0}=0.004$ km. Randomly generated
shadowing factors are used to associate mobiles with cell sectors and in other
computations. The code rate permitted by the threshold is given by%
\begin{equation}
R=\log_{2}\left(  1+l_{s}\beta\right)
\end{equation}
where $l_{s}=0.794$ corresponds to a 1 dB loss relative to the Shannon bound
for complex discrete-time AWGN channels. The density of mobiles in a network
of area $A_{\mathsf{net}}$ is $\lambda=M/A_{\mathsf{net}}.$ The outage
probability $\epsilon_{i}$ of the reference link for simulation trial $i$ is
computed by applying (\ref{a1})$-($\ref{a4}). The \emph{throughput} of the
reference uplink for simulation trial $i$ is $R(1-\epsilon_{i})$. The average
outage probability over all the simulation trials is
\begin{equation}
\overline{\epsilon}=\frac{1}{N}\sum\limits_{i=1}^{N}\epsilon_{i}.
\end{equation}
The maximum rate of successful data transmissions per unit area is
characterized by the \emph{area spectral efficiency}, defined as
\begin{equation}
\mathcal{A}=\lambda R\left(  1-\overline{\epsilon}\right)  \label{a8}%
\end{equation}
where the units are bits per channel use per unit area.

To avoid edge effects, the performance is measured for mobile stations placed
in the yellow shaded area of Fig. 1, which is a 20 km by 20 km square located
in the middle of the network. To consider the effect of base-station
densification, the base-station deployment and the density of mobiles
$\lambda=20/km^{2}$ are maintained while the size of the network is scaled
(dimensions redefined). For a fixed number of base stations, each scaling
reduces the associated number of mobiles and decreases the transmission distances
accordingly. In considering the effects of intersector interference, only the
strongest 30 signals were used, as the attenuation of further signals was
severe enough to make their individual effects negligible.

The slot duration is $T=0.5$ ms, $P_{r}/\mathcal{N}=30$ dB, $L/L_{j}%
=L/L_{l}=10,$ and $\mu=40/km$. Each base station has $b=0.01$ and$\ \zeta=24$.
Each mobile beamwidth is $\Theta=0.1\pi$ radians and sidelobe level is
$a=0.1$, and the common activity factor is $p_{i}=1$. The propagation
parameters are $\left(  m_{\max},m_{\min},\alpha_{\min},\alpha_{\max}%
,\sigma_{\min},\sigma_{\max}\right)  =$ $\ \left(  2,1,2.3,4.7,6.1\text{
dB},12.6\text{ dB}\right)$, which are suitable for an urban network. The reference signal has fading parameter
$m_{0}=1$ when $d_{r}=0.1$ km, and $m_{0}=2$ when $d_{r}=0.01$ km. The ratio
$C/M$ serves as a measure of the densification of the base stations.

To capture the impact of densification, we need to take into account the
decrease in the \emph{typical link-length }$d_{r}$ of the reference link and
the consequent increase in $\Gamma_{0}$ as C/M increases. For a fixed density
of mobiles, the average cell radius and hence a \emph{typical} $d_{r}$ are
proportional to $1/\sqrt{C/M}.$ For the range of C/M of interest, we consider
\begin{equation}
d_{r}=\frac{d_{r0}}{10\sqrt{C/M}},\text{\ \ }0.01\leq C/M\leq1\label{L}%
\end{equation}
where $d_{r0}$ is the typical link length when $C/M=0.01$.%

\begin{figure}[tb]%
\centering
\includegraphics[width=9cm]{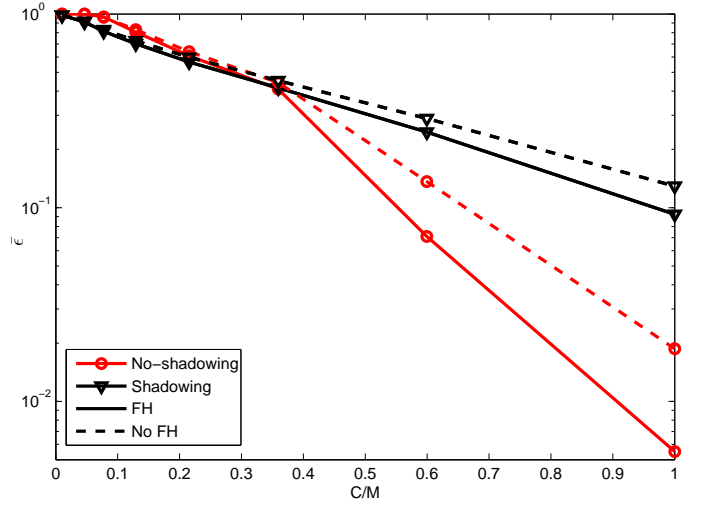}
\caption{Average outage probability for $d_{r0}=0.1$ km, $\beta=3$ dB,
$N=10^{5}$, and distance-dependent fading.} \label{Fig.3}
\end{figure}

\begin{figure}[tb]%
\centering
\includegraphics[width=9cm]{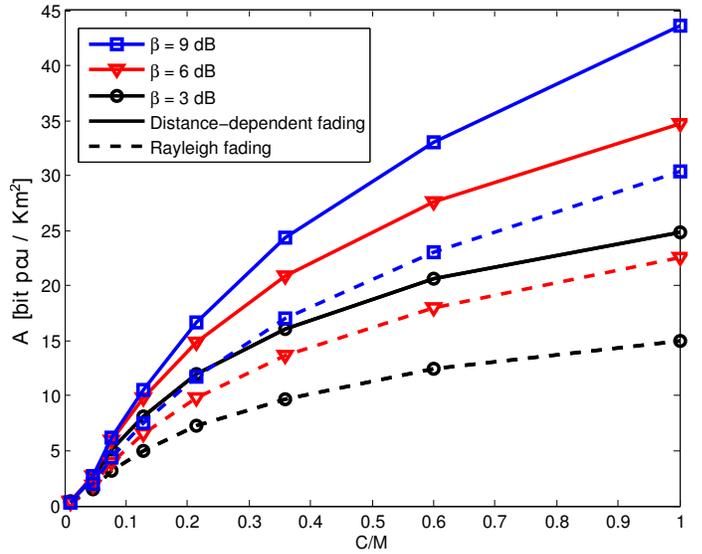}
\caption{Area spectral efficiency for three values of the \ac{SINR} threshold,
$d_{r0}=0.1$ km, and $N=10^{5}.$} \label{Fig.4}
\end{figure}

Fig. \ref{Fig.3} depicts the average outage probability $\overline{\epsilon}$ for a
typical link as a function of the densification for $d_{r0}=0.1$ km, $\beta=3$
dB, $N=10^{5}$, distance-dependent fading, both shadowing and no shadowing
($\sigma_{\min}=\sigma_{\max}=0)$, and both frequency hopping and its absence.
In the absence of frequency hopping, the reference signal is assumed to
experience constant fading over a subframe, which is valid if subframe
duration is less than the coherence time. The figure illustrates the reduced
importance of frequency hopping in the presence of substantial shadowing
because of the decrease in the typical values of $\Gamma_{0}.$ The primary
reasons for the monotonic decrease in $\overline{\epsilon}$ with densification
and the increased importance of the frequency hopping are the increases in
$\Gamma_{0}$ and the reduced shadowing and fading experienced by the reference
signal. Calculations show that decreasing $L/L_{j}$ has a very small
detrimental effect, which indicates the minor importance of the intercell
interference due to the sectorization and the narrow adaptive beams.

Fig. \ref{Fig.4} depicts the area spectral efficiency $\mathcal{A}$ as a function of
the densification for three values of the \ac{SINR} threshold, $d_{r0}=0.1$ km, and
$N=10^{5}$. Results are shown for both the distance-dependent fading and the
more severe Rayleigh fading for which $m_{i,j}=m(d)=m_{0}=1$. Increases in the
\ac{SINR} threshold $\beta$ of the network links increase the outage probability.
However, for sufficiently large values of C/M and hence $\Gamma_{0},$ this
effect is minor compared with increased code rate that can be accommodated. As
a result, the area spectral efficiency increases significantly.

\begin{figure}[tb]%
\centering
\includegraphics[width=9cm]{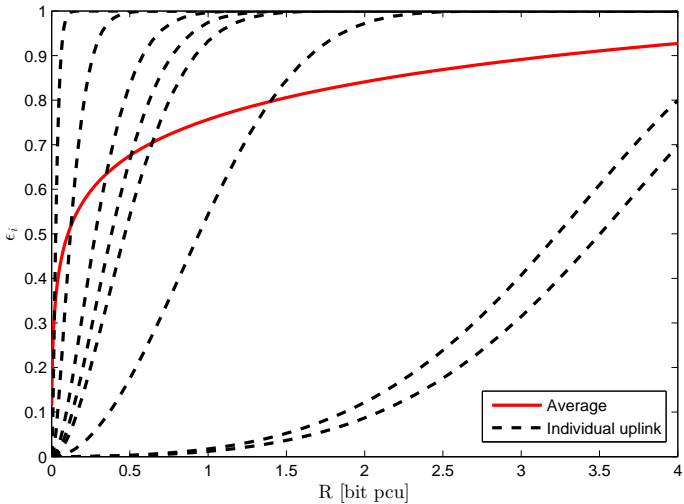}
\caption{Outage probability for eight uplinks and the average outage
probability over all the uplinks for $C/M=0.1,$ $\beta=3$ dB,
distance-dependent fading, and a single simulation trial.} \label{Fig.5}
\end{figure}

Fig. \ref{Fig.5} shows the outage probability as a function of code-rate $R$ for
$C/M=0.1,$ $\beta=3$ dB, distance-dependent fading, and a single simulation
trial. The dashed lines in the figure were generated by selecting eight random
uplinks in the network and computing each outage probability. The average over
all the uplinks is represented by the solid line. Despite the use of power
control, there is considerable variability in the dependence of the outage
probability on the code rate due to the irregular network topology.

\section{Conclusions}

This paper derives an analytical model for calculating the outage probability
and area spectral efficiency. The model includes the effects of
millimeter-wave propagation, directional beams, frequency hopping, an
arbitrary network topology, and the assignment of frequency blocks to mobiles.
Numerical examples illustrate the effects of various features and parameters.
Base-station densification improves the network performance significantly. The
significance of the intercell interference is greatly reduced because of the
highly directional sectorization and beamforming. The frequency hopping more
effectively compensates for frequency-selective fading as the shadowing
decreases. The area spectral efficiency increases with the code rate despite
the increase in the outage probability.

\bibliographystyle{ieeetr}
\bibliography{5Grefs}

\balance
\vfill

\end{document}